\documentclass[secnumarabic,aps,prd,floatfix,nofootinbib,showkeys,twocolumn,showpacs,10pt,superscriptaddress]
{revtex4-1}

\usepackage{times}
\usepackage{amsfonts}
\usepackage{amsmath}
\usepackage{amssymb}
\usepackage{natbib}
\usepackage{graphicx}
\usepackage{enumitem}
\usepackage{xcolor}
\usepackage[colorlinks=true,linkcolor=blue]{hyperref}
\usepackage[outdir=./]{epstopdf}
\usepackage[normalem]{ulem}

\def\aap{A\&A}
\def\aj{AJ}
\def\apj{ApJ}
\def\apjl{ApJL}
\def\apjs{ApJS}
\def\apss{Astrophysics and Space Science}
\def\jcap{JCAP}
\def\mnras{MNRAS}

\def\pasp{PASP}

\begin{document}

\title{{Energy transfer from baryons to dark matter as a unified solution to small-scale structure issues of the 
\texorpdfstring{$\Lambda$CDM}{LCDM} model}}

\author{A.~\surname{Del Popolo}}%
\affiliation{%
Dipartimento di Fisica e Astronomia, University Of Catania, Viale Andrea Doria 6, 95125, Catania, Italy
}
\affiliation{%
Institute of Modern Physics, Chinese Academy of Sciences, Post Office Box 31, Lanzhou 730000, Peoples Republic of China
}
\affiliation{%
INFN sezione di Catania, Via S. Sofia 64, I-95123 Catania, Italy
}
\email[Corresponding author. ]{adelpopolo@oact.inaf.it}

\author{Francesco~\surname{Pace}}
\affiliation{%
Jodrell Bank Centre for Astrophysics, School of Physics and Astronomy, The University of Manchester, Manchester, 
M13 9PL, United Kingdom
}
\email{francesco.pace@manchester.ac.uk}

\author{Morgan~\surname{Le~Delliou}}
\affiliation{%
Institute of Theoretical Physics, Physics Department, Lanzhou University, No.222, South Tianshui Road, Lanzhou, Gansu,  
730000, Peoples Republic of China}
\affiliation{%
Instituto de Astrof{\'i}sica e Ci{\^e}ncias do Espa{\c c}o, Universidade de Lisboa, Faculdade de Ci{\^e}ncias, 
Ed. C8, Campo Grande, 1769-016 Lisboa, Portugal}
\email{delliou@ift.unesp.br}

\author{Xiguo~\surname{Lee}}
\affiliation{%
Institute of Modern Physics, Chinese Academy of Sciences, Post Office Box 31, Lanzhou, Gansu 730000, Peoples Republic 
of China
}
\email{xgl@impcas.ac.cn}

\label{firstpage}

\date{\today}

\begin{abstract}
Using a semianalytic code, we show how baryon physics in a $\Lambda$CDM cosmology could solve the discrepancy between 
numerical predictions of dark matter haloes and observations, ranging from dwarf galaxies to clusters, without the 
need of nonstandard dark matter models as advocated, for example, by 
[Kaplinghat {\it et al.}, \href{https://doi.org/10.1103/PhysRevLett.116.041302}{Phys. Rev. Lett. {\bf 116}, 041302, 
(2016)}]. Combining well established results, we show, for the first time, how accounting for baryon physics, in 
particular dynamical friction mechanisms, leads to flat galaxy-cluster profiles and correlations in several of their 
properties, solves the so-called ``diversity problem'' and reproduces very well the challenging, extremely low-rising 
rotation curve of IC2574. We therefore suggest treating baryonic physics properly before introducing new exotic 
features, albeit legitimate, in the standard cosmological model.
\end{abstract}



\maketitle

\section{Introduction}
The $\Lambda$CDM model, while very successful \citep{Spergel2003,Komatsu2011}, presents some issues 
(e.g., \cite{Weinberg1989,Astashenok2012}). 
Particularly troublesome is the discrepancy between the flat density profiles of dark-matter (DM)-dominated dwarf 
galaxies, irregulars and low surface brightness galaxies (hereafter LSBs), high surface brightness spiral galaxies, in 
some cluster of galaxies, and the cuspy profile predicted by dissipationless N-body simulations 
(e.g., \cite{Navarro1996}), dubbed cusp/core problem \citep{Moore1994,Flores1994,Cardone2012}. 
Better understood in terms of the excess of DM in the inner parts of the galaxies rather than of the inner slope, 
it connects to the too-big-to-fail (TBTF) problem \citep{BoylanKolchin2011,Papastergis2015}. 
We also mention the large diversity (hereafter dubbed ``diversity problem" in analogy to other works) in the 
dwarf-galaxies rotation curves (RCs), which are at odds with hydrodynamic simulations\footnote{Note that the 
$\Lambda$CDM model suffers from others drawbacks like the cosmological constant problem 
\cite{Weinberg1989,Astashenok2012}, and the cosmic coincidence problem.}.

A possible solution to these problems is to assume that the DM component is not cold and this leads to a wealth of 
different models (e.g., self-interacting DM (SIDM) models \cite{Spergel2000}) recently used by \cite{Kaplinghat2016} to 
propose a unified solution (at all scales) to the small scale problems of the $\Lambda$CDM model.

The study of \cite{Kaplinghat2016} claims the difficulty or impossibility for the $\Lambda$CDM model to explain the 
rotation curve of IC2574 and several other issues, and, in an attempt to solve them, proposes the SIDM model as a 
possible alternative to the $\Lambda$CDM model.

It is thus of fundamental importance to verify whether the $\Lambda$CDM model can solve, at all scales, the problems 
discussed in \cite{Kaplinghat2016}.

In the present paper we want to address the following question: does a unified solution exist to the ``deficit 
problem in halos" in the $\Lambda$CDM model without invoking a different physics? 
We will closely follow \cite{Kaplinghat2016} in showing how the ``deficit problem in halos" is solved when baryonic 
physics is taken into account in the $\Lambda$CDM model.

The plan of this work is as follows: we first present our semianalytical model in Sec.~{\ref{sect:model}}. 
Differently from N-body and hydrodynamical simulations, the model presents the various physical contributions more 
clearly and allows to disentangle them more easily. We then show, in Sec.~{\ref{sect:clusters}}, how it reproduces the 
clusters (e.g., A2537) presented in \cite{Newman2013a,Newman2013b}, how it explains the RCs of some peculiar galaxies 
(e.g., IC 2574, this one fitted by \cite{Creasey2017} using SIDM and baryon physics) in Sec.~{\ref{sect:dwarf}}, and 
how it solves the diversity problem discussed by \cite{Oman2015} in Sec.~{\ref{sect:diversity}}. 
Finally we conclude in Sec.~{\ref{sect:conclusions}} discussing our results in comparison to those of 
\cite{Kaplinghat2016}.

\section{Model}\label{sect:model}
Here we recall the model used above. First introduced by \cite{DelPopolo2009,DelPopolo2009a}, many of its features were 
developed in \cite{DelPopolo2014a,DelPopolo2016a,DelPopolo2016b}, and allowed studies of the density profiles' 
universality \citep{DelPopolo2010,DelPopolo2011}, of the density profile of galaxies 
\citep{DelPopolo2012a,DelPopolo2014,DelPopolo2013} and clusters 
\citep{DelPopolo2012b,DelPopolo2014}, and of the inner surface-density of galaxies \citep{DelPopolo2013d}.

The model is a strong improvement of the original spherical collapse model
\citep{Gunn1972,Bertschinger1985,Hoffman1985,Ryden1987,Ascasibar2004,Williams2004}, and includes the effects of 
random angular momentum due to random motions arising in the collapse phase (e.g., \citep{Ryden1987,Williams2004}) and 
ordered angular momentum (e.g., \citep{Ryden1988,DelPopolo1997,DelPopolo2000}) from tidal torques, dark energy 
\citep{DelPopolo2013a,DelPopolo2013b,DelPopolo2013c,Pace2014b}, dynamical friction arising from the interaction between 
baryons and DM 
\citep{ElZant2001,ElZant2004,Ma2004,RomanoDiaz2008,RomanoDiaz2009,DelPopolo2009,Cole2011,Inoue2011,Nipoti2015},
adiabatic contraction (e.g., \citep{Blumenthal1986,Gnedin2004, Klypin2002,Gustafsson2006}), gas cooling, star 
formation, photoionization, supernovae and AGN feedback \citep{DeLucia2008,Li2010,Martizzi2012}.

At this stage we want to stress that the main mechanism of the model resides in the dynamical friction and as described 
in point \ref{SNmech} (see the following), the other mechanisms (e.g. SN feedback) are only contributing at the percent 
level.

The main evolutionary phases present in the model are listed as follows:
\begin{enumerate}
\item In the linear phase the DM and the diffuse gas proto-structure expand to a maximum radius, then DM 
      re-collapses, forming the potential well for baryons to fall into.
\item Baryons are subject to radiative processes and form stars, condense forming clumps which collapse in 
      the centre of the halo.
\item While baryons are compressed during the collapse phase and make the DM profile more cuspy, the formed clumps 
      interact with the DM component and stars via dynamical friction (DF) and transfer energy and angular momentum
      (AM) \citep{Read2005,Pontzen2012,Teyssier2013} with the result of having DM particles moving away from the centre 
      leading to a net reduction of the central density \citep{ElZant2001,ElZant2004}. 
      This process leads to the formation of cores in dwarf spheroidals and spirals, while giant galaxies keep a 
      steeper profile due to their deeper potential wells.
\item The effect of DF is amplified by tidal torques and random AM.
\item \label{SNmech} Finally, in a later phase, supernovae explosions expel gas lowering the stellar 
      density and disrupting smallest gas clumps after their partial conversion into stars (see \cite{Nipoti2015}).  
      The expulsion of gas has the additional effect of enlarging the core (this is quantified in a few percent effect).
\end{enumerate}

\subsection{Density profile}
The spherical density perturbation model follows the perturbation evolution from the expansion due to the Hubble 
flow till the turn-around and then into collapse \citep{Gunn1977,Fillmore1984}. 
The final profile of the perturbation is given by
\begin{equation}\label{eq:dturnnn}
 \rho(x)=\frac{\rho_{\rm ta}(x_{\rm m})}{f(x_{\rm i})^3}
 \left[1+\frac{d\ln{f(x_{\rm i})}}{d\ln{g(x_{\rm i})}}\right]^{-1}\;,
\end{equation}
where $x_{\rm i}$ is the initial radius, $f(x_{\rm i})=x/x_{\rm m}(x_{\rm i})$ the collapse factor, 
$\rho_{\rm ta}(x_{\rm m})$ the turn-around density and $x_{\rm m}(x_{\rm i})$ the turn-around radius, which can be 
written as
\begin{equation}
 x_{\rm m} = g(x_{\rm i}) 
           = x_{\rm i}\frac{1+\bar{\delta}_{\rm i}}{\bar{\delta}_{\rm i}-(\Omega_{\rm i}^{-1}-1)}\;.
\end{equation}
In the above expression, $\Omega_{\rm i}$ represents the matter density parameter and $\bar{\delta}_{\rm i}$ the 
average overdensity inside a shell of DM and baryons. Initially, in the gas phase, the ``universal baryon fraction" is 
set to $f_{\rm b}=0.17\pm 0.01$ (0.167 in \cite{Komatsu2011,Komatsu2009}) and evolves, due to star formation 
processes, as follows.

Using the tidal torque theory (TTT), one can evaluate the ``ordered angular momentum" $h$ induced by tidal torques 
exerted by large scale structures on smaller scales 
\citep{Hoyle1953,Peebles1969,White1984,Ryden1988,Eisenstein1995,Hiotelis2013}. 
At the same time, the ``random angular momentum" $j$ depends on the eccentricity $e=r_{\rm min}/r_{\rm max}$ 
\citep{AvilaReese1998}, being $r_{\rm max}$ the apocentric radius, and $r_{\rm min}$ the pericentric radius. One needs 
to correct the eccentricity for the effects of the dynamical state of the system found by \cite{Ascasibar2004}:
\begin{equation}
 e(r_{\rm max})\simeq 0.8\left(\frac{r_{\rm max}}{r_{\rm ta}}\right)^{0.1}\;,
\end{equation}
for $r_{\rm max}<0.1 r_{\rm ta}$, being $r_{\rm ta}=x_{\rm m}(x_{\rm i})$ the spherically averaged turn-around 
radius.

The steepening of the profile due to adiabatic compression is calculated iteratively \cite{Spedicato2003} and follows 
the prescription of \cite{Gnedin2004}, while dynamical friction is modelled by a force in the equation of motions (see 
\cite{DelPopolo2009}, Eq.~(A14)).

\subsection{Baryons, discs, and clumps}
Baryons are initially modelled in a gas phase and settle into a rotationally supported stable disk for spiral galaxies. 
The disc sizes and masses obtained from the model have been shown to solve the angular momentum catastrophe (AMC) 
(Sect. 3.2, Figs. 3, and 4 of \cite{DelPopolo2014}), producing discs with sizes and masses similar to those of real 
galaxies.

However, denser discs are unstable, as known from the Jeans' criterion, despite the shear force stabilisation. Toomre 
\cite{Toomre1964} gave the disc instability and clump formation condition

\begin{equation}
 Q \simeq \frac{\sigma \Omega}{\pi G \Sigma} = \frac{c_{\rm s} \kappa}{\pi G \Sigma}<1\;,
\end{equation}
with $\sigma$ the 1-D velocity dispersion,\footnote{In most galaxies hosting clumps $\sigma\simeq 20-80$~km/s.} 
$\Omega$ the angular velocity, $\Sigma$ the surface density, $c_{\rm s}$ the adiabatic sound speed, and $\kappa$ the 
epicyclic frequency. 
The fastest growing mode derives from the perturbation dispersion relation when $Q<1$, and is the solution of 
$d\omega^2/dk=0$, giving $k_{\rm inst}=\frac{\pi G \Sigma}{c_s^2}$ (see \cite{BinneyTremaine1987,Nipoti2015}).
We used that condition to obtain the clumps radii for our galaxy \citep{Krumholz2010}
\begin{equation}
 R \simeq 7 G \Sigma/\Omega^2 \simeq 1 {\rm kpc}\;.
\end{equation}
Marginally unstable discs ($Q \simeq 1$) with maximal velocity dispersion have a total mass three times larger than 
that of the cold disc, form clumps with mass $\simeq 10\%~M_{\rm d}$ \citep{Dekel2009}, where $M_{\rm d}$ is the mass 
of the disk.

Objects of masses few times $10^{10}~M_{\odot}$, and found in $5 \times 10^{11} M_{\odot}$ haloes at $z \simeq 2$, stay 
in a marginally unstable phase for $\simeq 1$~Gyr. In general we found that the main properties of clumps are similar 
to those found by \cite{Ceverino2012}.

In agreement with 
\cite{Ma2004,Nipoti2004,RomanoDiaz2008,RomanoDiaz2009,DelPopolo2009,Cole2011,Inoue2011,DelPopolo2014d,Nipoti2015}, 
energy and AM transfer from clumps to DM flatten the profile more efficiently in smaller haloes.

\subsubsection{Clumps lifetime}
The clumps discussed above have been observed both in simulations (e.g., 
\cite{Ceverino2010,Perez2013,Bournaud2014,Ceverino2014,Perret2014,Ceverino2015,Behrendt2016}) and observations.

Clumpy structures, dubbed chain galaxies, and clump clusters are observed in high-redshift galaxies 
\citep{Elmegreen2004,Elmegreen2009,Genzel2011}, while massive star-forming clumps \citep{Guo2012,Wuyts2013} 
were found in HST Ultra Deep Field observations, in a large number of star-forming $z=1-3$ galaxies \citep{Guo2015}, 
and even in deeper fields up to $z \simeq 6$ \cite{Elmegreen2007}.

Very gas-rich disc experiencing radiative cooling in the dense accreting gas induces self-gravity instability 
that should form those clumpy structures (e.g., \citep{Noguchi1998,Noguchi1999,Aumer2010,Ceverino2010,Ceverino2012}).

The clump's lifetime is crucial in the processes described above: they should transform the cusp into a core, before 
disruption by stellar feedback, by sinking to the galaxy centre. The mass fraction lost by stellar feedback, $e$, 
and that transformed into stars $\varepsilon=1-e$, assess the capacity of the clump to form a bound stellar system. 
According to simulations and analytic models \cite{Baumgardt2007}, for $\epsilon \geq 0.5$, most of the stellar mass 
will remain bound. Estimation by \citep{Krumholz2010} of the expulsion fraction 
$e=1-\varepsilon=0.86 (\Sigma_1 M_9)^{-1/4} \epsilon_{\rm eff},_{-2}$, with 
$\Sigma_1=\frac{\Sigma}{0.1~{\rm g/cm}^2}$, $M_9=M/10^9 M_{\odot}$, and $\epsilon_{\rm eff},_{-2}=\epsilon_{eff}/0.01$, 
can be obtained using the dimensionless star-formation rate efficiency 
$\epsilon_{\rm eff}=\frac{\dot{M}_{\ast}}{M/t_{\rm ff}}$, i.e. the free-fall $t_{\rm ff}$ to depletion time ratio, with 
stellar mass $M_{\ast}$. For a broad range of densities, size scales and environments, $\epsilon_{eff} \simeq 0.01$ 
\citep{Krumholz2007}. For a typical clump with $M \simeq 10^9 M_{\odot}$, $\Sigma \simeq 0.1~{\rm g/cm}^2$ and 
$\epsilon_{eff} \simeq 0.01$ one finds$\varepsilon \simeq 0.85$ and $e \simeq 0.15$. Thus clumps mass loss is 
small before they reach the galactic centre. 
The previous argument and the relation $e=1-\varepsilon$ are valid for smaller clumps, which are however more 
compact and the corresponding galaxies smaller, with the final result that clumps reach the centre before being 
destroyed.

An alternative assessment of clumps disruption compares its lifetime to its central migration time. 
Dynamical friction and tidal torques are responsible for a migration time of $\simeq 200$~Myrs for a $10^9 M_{\odot}$ 
mass clump, \citep{Genzel2011,Nipoti2015}.

For clumps similar to those discussed above, the Sedov-Taylor solution (\citep{Genzel2011}, Eqs. 8,9) yields an 
expansion timescale similar to the migration time.

Ceverino {\it et~al.} \citep{Ceverino2010} found that in their hydro-dynamical simulations clumps are in Jeans' 
equilibrium and rotationally supported and therefore a long lifetime ($\simeq 2 \times 10^8$~Myr) can be inferred. 
These findings are in agreement with \citep{Krumholz2010} who found that clumps survive if only few percent of the gas 
is converted into stars, in agreement with the Kennicutt-Schmidt law. All these results confirm previous simulations by 
\cite{Elmegreen2008}.

While \citep{Bournaud2014,Ceverino2014,Perret2014} found long-living clumps reaching the centre when properly taking 
into account stellar feedback, radiative and non-thermal feedback, \cite{Perez2013} obtained long-lived objects for any 
reasonable amount of feedback.

Strong observational evidence of long-lived clumps derives from the estimation of clump ages through gas expulsion, 
metal enrichment and expansion time scales (respectively 170-1600~Myrs, $\simeq 200$~Myrs and $>100$~Myrs) 
by \cite{Genzel2011}.

Finally, clump stability is supported by observational similitude in radius and mass between low- and high-redshift 
clumps \citep{Elmegreen2013,Garland2015,Mandelker2017}.

\subsection{Star formation and feedback}
The model treatment of gas cooling, star formation, reionisation and supernovae feedback follows the lines of 
argument of \citep{DeLucia2008,Li2010}.

Reionisation occurs in the redshift range 11.5-15, changing the baryon fraction as \citep{Li2010}
\begin{equation}
 f_{\rm b, halo}(z,M_{\rm vir}) = \frac{f_{\rm b}}{[1+0.26 M_{\rm F}(z)/M_{\rm vir}]^3}\;,
\end{equation}
with $M_{\rm vir}$ the virial mass and $M_{\rm F}$ the ``filtering mass" (see \cite{Kravtsov2004}), while a 
cooling flow models the gas cooling (e.g., \citep{White1991,Li2010}, see Sect. 2.2.2).

Star formation occurs after gas settles in a disk, at a rate given by
\begin{equation}
 \psi = 0.03 M_{\rm sf}/t_{\rm dyn}\;,
\end{equation}
resulting in the gas mass conversion into stars 
\begin{equation}
 \Delta M_{\ast} = \psi\Delta t\;.
\end{equation}
Here, we denote with $M_{\rm sf}$ the gas mass above the density threshold $n>9.3/{\rm cm^3}$ 
(fixed as in \cite{DiCintio2014}), and we use the time-step $\Delta t$ for a disc dynamical time $t_{\rm dyn}$ 
(see \cite{DeLucia2008}, for more details).

Following \cite{Croton2006}, supernovae feedback (SNF) injects energy for SN explosions as
\begin{equation}
 \Delta E_{\rm SN}=0.5\epsilon_{\rm halo}\Delta M_{\ast} \eta_{\rm SN}E_{\rm SN}\;, 
\end{equation}
modelling, for a Chabrier IMF \cite{Chabrier2003}, the number of supernovae per solar mass as 
$\eta_{\rm SN}=8\times 10^{-3}/M_{\odot}$, the disc gas reheating energy efficiency $\epsilon_{\rm halo}$, and the 
typical SN explosion energy release as $E_{\rm SN}=10^{51}$ erg.

The resulting gas reheating depends on the stars formed
\begin{equation}
 \Delta M_{\rm reheat} = 3.5 \Delta M_{\ast}\;.
\end{equation}

The thermal energy change caused by the reheated gas 
\begin{equation}
\Delta E_{\rm hot} = 0.5\Delta M_{\rm reheat} \eta_{\rm SN}E_{\rm SN}\;,
\end{equation}
induces, if $\Delta E_{\rm SN}>\Delta E_{\rm hot}$, ejection of 
\begin{equation}
 \Delta M_{\rm eject} = \frac{\Delta E_{\rm SN}-\Delta E_{\rm hot}}{0.5 V^2_{\rm vir}}\;,
\end{equation}
hot gas from the halo, with $V^2_{\rm vir}=2\Delta E_{\rm hot}/\Delta M_{\rm reheat}$.

A fundamental difference between the SNF model by \cite{DiCintio2014} and ours resides in our cusp flattening starting 
before star formation, and its energy source being gravitational.

Stellar and supernovae feedback start to take place only after DF formed the core, disrupting the core gas clouds 
(similarly to \cite{Nipoti2015}).

AGN quenching becomes significant for masses $M\simeq 6 \times 10^{11} M_{\odot}$ \citep{Cattaneo2006}. 
The prescription of \cite{Martizzi2012,Martizzi2012a} leads to our account for AGN feedback: when stellar density 
exceeds $2.4 \times 10^6 M_{\odot}/{\rm kpc}^3$, gas density reaches 10 times that, and the 3D velocity dispersion 
exceeds 100~km/s, a Super-Massive-Black-Hole (SMBH) is formed, seeded at $10^5~M_{\odot}$. Hence, a modification of 
the model by \cite{Booth2009}, as in \cite{Martizzi2012}, yields SMBH mass accretion and AGN feedback.

Our model demonstrated its robustness in several ways:   
\renewcommand{\theenumi}{\alph{enumi}}
\begin{enumerate}
 \item The cusp flattening from DM heating by collapsing baryonic clumps predicted for galaxies and clusters is 
 in agreement with following studies 
 \citep{ElZant2001,ElZant2004,RomanoDiaz2008,RomanoDiaz2009,Cole2011,Inoue2011,Nipoti2015}; \citep{DelPopolo2011}, 
 in Fig.~4, shows a comparison of our model with the SPH simulations of \cite{Governato2010}.
 \item The correct shape of galaxies density profiles \citep{DelPopolo2009,DelPopolo2009a} was predicted before 
 the SPH simulations by \cite{Governato2010,Governato2012}, and correct clusters density profiles 
 \citep{DelPopolo2012b} before \cite{Martizzi2013}, although the latter use different dominant mechanisms than the 
 former.
 \item The inner slope dependence on halo mass (\cite{DelPopolo2010}, Fig. 2a solid line)was predicted before 
 almost the same result was shown in Fig. 6 (the non-extrapolated part of the plot) of \cite{DiCintio2014}, 
 in terms of $V_{\rm c}$ (which is $2.8 \times 10^-2 M_{\rm vir}^{0.316}$ \citep{Klypin2011}).\\
 We also found that the inner slope depends on the total baryonic content to total mass ratio \cite{DelPopolo2012b}, 
 as seen later in \cite{DiCintio2014}.\\
 \citep{DelPopolo2016a,DelPopolo2016b} show a comparison of the change of the inner slope with mass with the 
 simulations of \citep{DiCintio2014}. Moreover Figs.~4 and 5 in \citep{DelPopolo2016a,DelPopolo2016b}, 
 show a comparison of the Tully-Fisher, Faber-Jackson, $M_{\rm Star}-M_{\rm halo}$ relationships with simulations.
\end{enumerate}

\begin{figure*}[t]
 \includegraphics[height=5cm,angle=0]{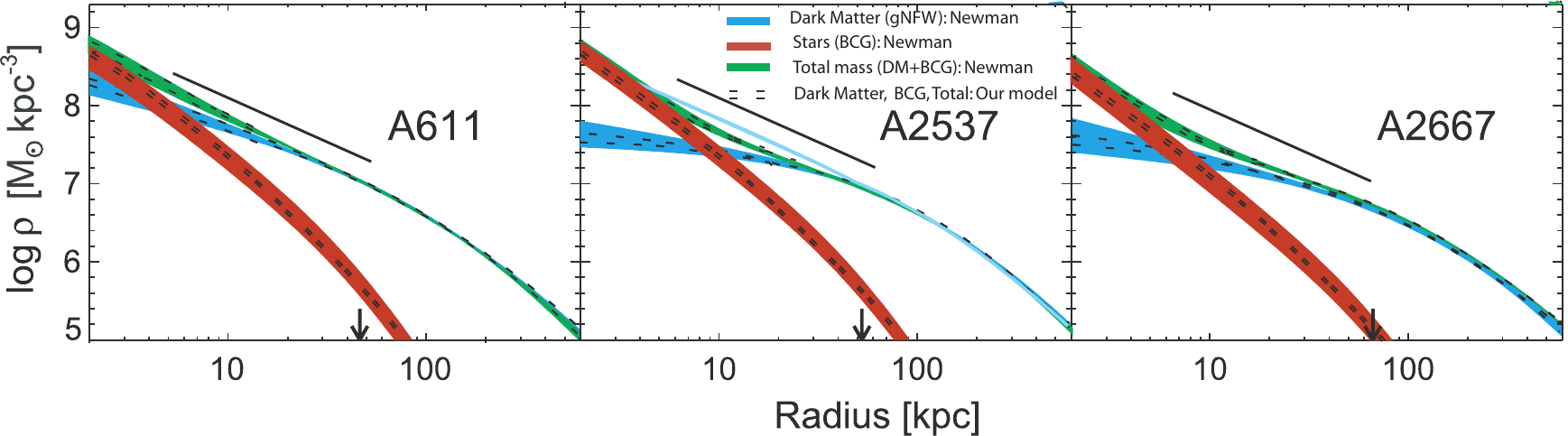}
 \caption[justified]{Density profile of the total and DM mass for the clusters A611 (left panel), A2537 (
 middle panel) and A2667 (right panel). The bottom blue (upper green) band represents the DM (total mass) 
 density profile determined by \cite{Newman2013a,Newman2013b}, while the red band the stars' mass. The band in black 
 dashed lines is the DM density profile obtained in this paper. The band widths represent the 1-$\sigma$ uncertainty. 
 The bottom arrow, in each panel, is the three-dimensional half-light radius of the BCG. 
 The segment with slope $r^{-1.13}$ spans the radial range $r=[0.003-0.03]r_{200}$. 
 The light blue line in the middle panel represents the NFW profile.}
 \label{fig:bgfr}
\end{figure*}

\section{Clusters}\label{sect:clusters}
The cusp-core problem extends to the scales of cluster of galaxies: combining weak and strong lensing and stellar 
kinematics, the total inner density profile was shown as well described by dissipationless N-body simulations at 
radii $>5-10$~kpc, while DM profiles are flatter than those obtained in the simulations 
\citep{Newman2013a,Newman2013b}, within a radius of $\simeq 30$~kpc, typical of the Brightest Cluster Galaxy (BCG) 
radius. The DM profile is characterised by a variation of the slope, $\alpha = -d \log \rho_{\rm DM}/d \log r$, from 
cluster to cluster which correlates with the BCG properties.

The total and the DM density profiles of MS2137, A963, A383, A611, A2537, A1667 and A2390, were determined in the 
aforementioned works. In \cite{Newman2013a,Newman2013b}, improved data allowed the determination of the stellar mass 
scale, allowing to produce a more physically consistent analysis, reducing the degeneracy among stellar and dark 
mass, and taking into account the BCGs homogeneity.

Two different approaches are available to compare the result of our model to the density profiles in 
\cite{Newman2013a}. 
While the density profile depends on the virial halo mass, $M_{\rm vir}$, on the baryon fraction, 
$f_{\rm b}=M_{\rm b}/M_{\rm vir}$, and on the random AM, as shown in \cite{DelPopolo2012b}, one can adjust the 
value of $j$, so that $\rho_{\rm DM}$ reproduces the observed clusters profiles. 
Here we rather prefer the possibility to ``simulate" the formation and evolution of clusters with similar 
characteristics to those of \cite{Newman2013a}: 
final halo mass,\footnote{We use $M_{200}$ as in \cite{Newman2013a}} $M_{\rm BCG}$, core radius, etc.,
\footnote{
By ``simulate" we mean that, as in hydrodynamic simulations, we fix the initial conditions and follow the evolution of 
galaxies, or clusters, from the linear to the non-linear phase, until the object formation, and its subsequent 
evolution due to the physical effects previously described.} 
by targeting not only the total halo and baryonic masses but also the observed radial density distribution (in the 
case of Fig.~\ref{fig:bgfr}) and rotation curve (in the case of Fig.~\ref{fig:RC}).

To obtain reasonable agreement between ours' and observed clusters, we run simulations till the final halo and baryonic 
masses and corresponding radial density distributions differ by at most 10\% from the observed clusters. To get this 
results the clusters were resimulated 50 times. 
To be precise, we iterated the model for each cluster 50 times until the produced clusters exhibited one with DM 
halo, gas, and stars masses and radial density profiles within 10\% of the observed target.

Of all the seven cited clusters, in Fig.~\ref{fig:bgfr}, we plot the spherically-averaged density profile of A2537  
(studied by \cite{Kaplinghat2016}), A611, and A2667 for the DM halo, BCG stars and their sum (total mass). 
The bottom blue band (upper green band) represents the DM (total mass) density profile determined by \cite{Newman2013a} 
for the clusters A611, A2537 (studied by \cite{Kaplinghat2016}), and A2667, fitting the profile with a generalized 
Navarro-Frenk-White (gNFW) profile. The band in black dotted lines is the DM density profile obtained with the model of 
this paper. The band width represents the 1-$\sigma$ uncertainty. The bottom arrow, in each panel, is the 
three-dimensional half-light radius of the BCG. The segment with slope $r^{-1.13}$ spans the radial range 
$r=[0.003-0.03]r_{200}$. The light blue line in the middle panel represents the NFW profile.

\begin{figure*}[!ht]
 \includegraphics[height=7cm,angle=0]{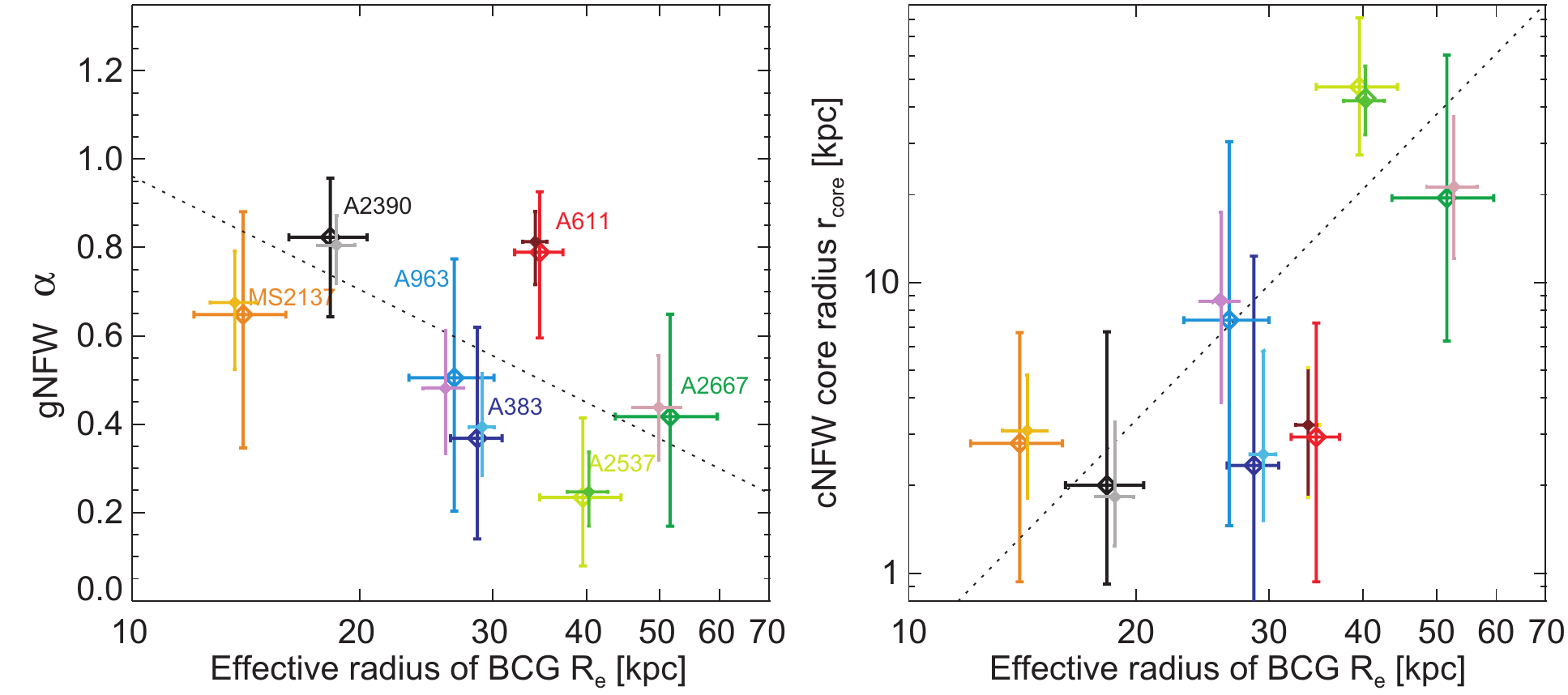}
 \caption{Correlation among the inner DM profile and the BCG size. 
 Left panel: inner slope of DM haloes, $\alpha$ vs the effective radius $R_{\rm e}$ of the BCG of the indicated 
 clusters obtained fitting the clusters density profile with the gNFW model. The larger (smaller) error-bars 
 correspond to the results by \citep{Newman2013b} (our model). The dotted lines are the least-square fits. 
 Right panel: core radii, $r_{\rm core}$, obtained fitting the clusters density profile with a cNFW model vs 
 $R_{\rm e}$.}
 \label{fig:corr}
\end{figure*}

The plot shows a good agreement between the observations and the model: an inner DM profile with almost flat slopes in 
the case of A2537, and much more cuspy as in A611. However, in all the sample the inner slope has an average of 
$\simeq 0.54$, flatter than the NFW profile; a total mass profile close to a NFW profile and baryons dominate the 
profile in the inner $\simeq 10$ kpc.

In the left and right panels of Fig.~\ref{fig:corr}, we show some correlations found by \citep{Newman2013b}. 
The points with error-bars in the middle panel show the value of the inner slope $\alpha$ vs the effective radius, 
$R_{\rm e}$, of the BCG for the clusters indicated, obtained fitting their profiles with the generalized 
Navarro-Frenk-White profile (gNFW). The larger error bars represents the result by \cite{Newman2013b}, the smaller 
ones, our results. The right panel represents the core radii, $r_{\rm core}$, vs $R_{\rm e}$ for the same clusters, 
obtained fitting the density profiles with a cored NFW model (cNFW) \cite{Newman2013b}. 
Dotted lines are the least-square fits. We point out that while the correlations observed by \citep{Newman2013b} 
are re-obtained in our model, simulations, usually reaching mass scales of $10^{11}-10^{12}~M_{\odot}$, do not. 
Even the simulations by \citep{Martizzi2012,Martizzi2013} observe a flattening of the inner profile but do not study 
the correlations.

\begin{figure}
 \includegraphics[height=5cm,angle=0]{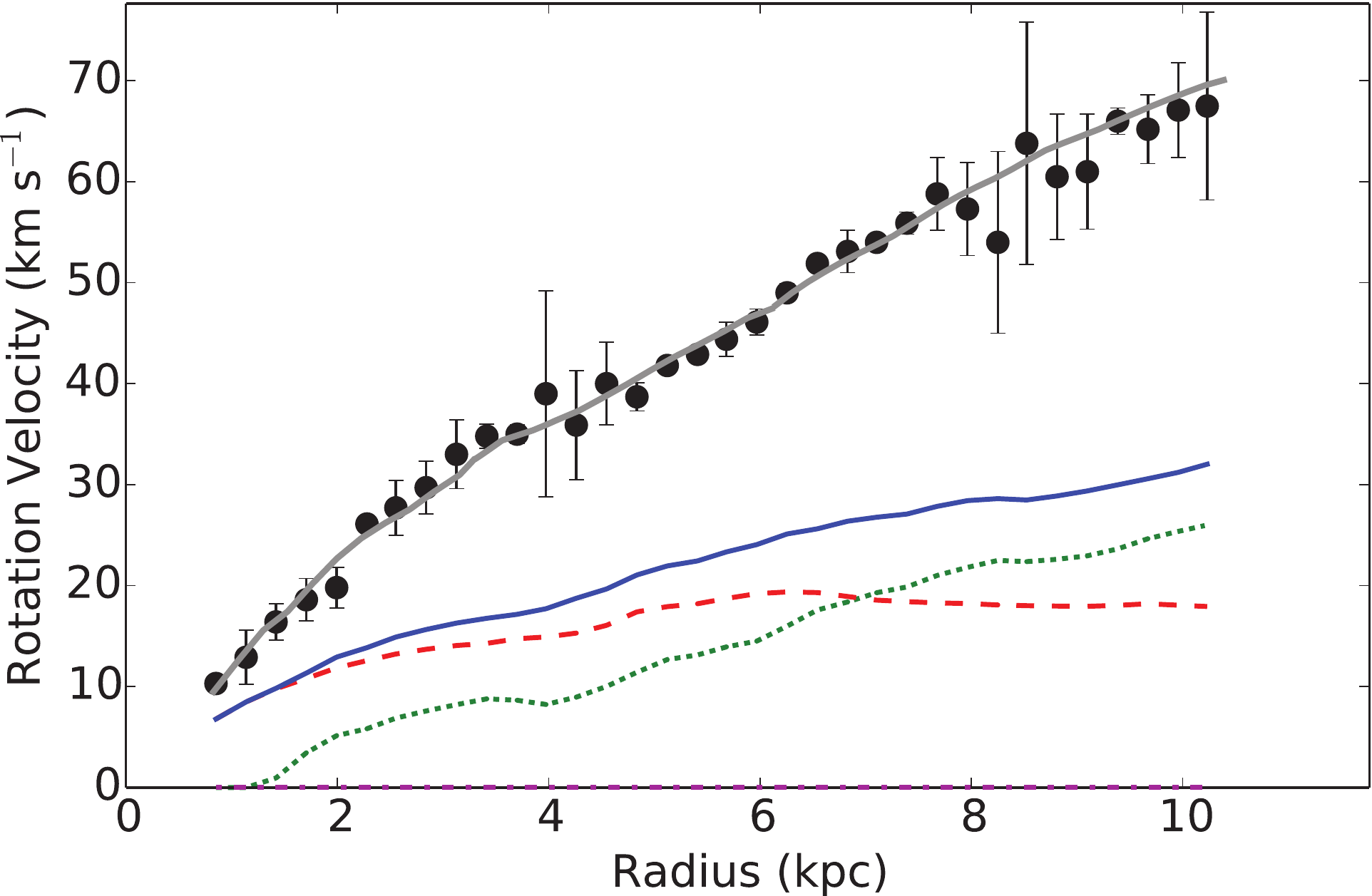}
 \caption[justified]{Rotation curve of IC2574, dot with error bars, from the SPARC catalogue. 
 The solid grey line is the RC of our model. The stars contribution to the RC is represented by the dotted green line, 
 the gas disk by the dashed red line, and the total baryonic mass by the continuous blue line. This RC is presented for 
 comparison with that shown by \cite{Kaplinghat2016}.}
 \label{fig:RC}
\end{figure}

\section{Dwarf and LSBs galaxies}\label{sect:dwarf}
We use our model to simulate 100 galaxies in a $\Lambda$CDM cosmology with similar characteristics to the SPARC 
sample \cite{Lelli2016}, a collection of nearby galaxies high-quality RCs. 
The stellar mass of the simulated sample is in the range $M_{\ast}=6 \times 10^6-10^{11}~M_{\odot}$. 
Of the galaxies used by \cite{Kaplinghat2016}, IC 2574, NGC 2366, DDO 154, UGC 4325, F563-V2, F563-1, F568-3, UGC 5750, 
F583-4, F583-1 are included in our larger sample.

\begin{figure*}[!ht]
 \includegraphics[height=7cm,angle=0]{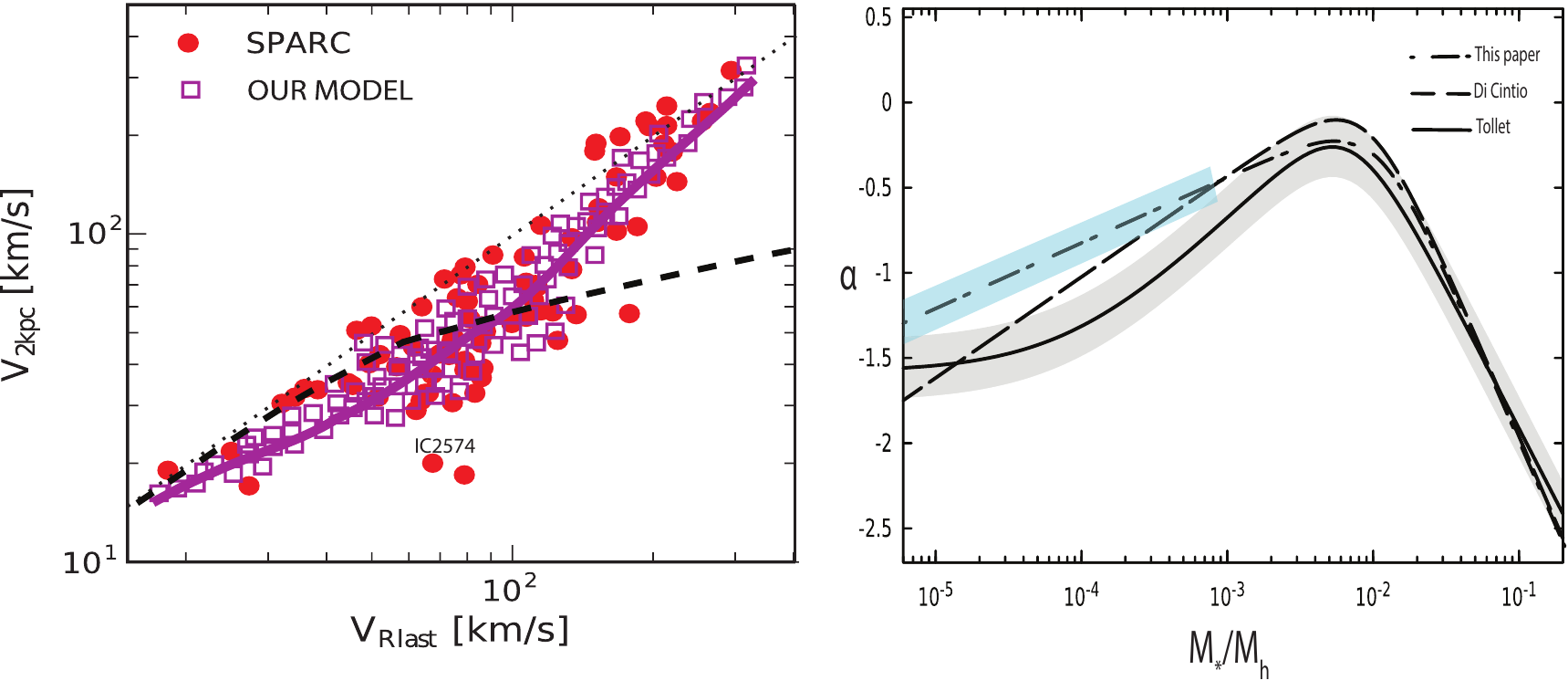}
 \caption[justified]{Left panel: Effects of baryonic physics on the relation $V_{\rm 2kpc}$-$V_{\rm Rlast}$. The 
 dashed line represents the expectation if the haloes were all described by a NFW profile. The dots come from 
 the SPARC galaxies of \cite{Lelli2016}, while the open squares represent the prediction of our model. Right panel: 
 Inner slope of the DM halo vs $M_{\ast}/M_{\rm halo}$. The dashed line is the result of \citep{DiCintio2014}, the 
 dot-dashed and the solid lines are the present paper and the results by \citep{Tollet2016}. The shaded blue and grey 
 regions represent the 1-$\sigma$ scatter in our result and \citep{Tollet2016}, respectively.}
\label{fig:VV}
\end{figure*}

The simulated objects produce a fair sample within the $\Lambda$CDM paradigm. The mass distribution follows the halo 
mass function. In particular we used \cite{DelPopolo2017}. The size distribution is a log-normal, as in 
(\cite{Shen2003}, Eq.~(12)), and comes from the log-normal distribution of the spin parameter. 
The observed RCs were compared to the most similar simulated galaxies (e.g., with same halo and baryonic mass). 
Of particular interest, IC2574 can be mistaken as an outlier if the error bars, of $\simeq 5-10~{\rm km/s}$, are 
not taken into account. However it lies in the outskirts of the distribution and we re-simulated it as we did 
previously for the clusters of our model to obtain an appropriately similar galaxy, within some percent, of the mass 
distributions of IC2574. Note that no further tuning was required. 
As an example, in Fig.~\ref{fig:RC}, we show the RC of IC2574, the same reproduced by SIDM in \cite{Kaplinghat2016}.

Similarly to the case of clusters, we run several simulations for this galaxy until the final halo and baryonic 
masses\footnote{Also, here, we refer not only to the total halo and baryonic masses but also the observed radial 
density distribution (RC).} differ by at most 10\% from the observational data. 
The plot shows the observational RC (dots with error-bars), while the model prediction for the RC, the contribution to 
the RC given by gas, stars, and the total baryonic mass, are represented by the grey continuous line, red dashed line, 
green dotted line, and blue continuous line, respectively.

As can be seen here, the RC of IC2574 (advocated to be problematic for the scenario of core formation \cite{Oman2015}) 
is very well described by the simulated galaxy, as well as the baryonic mass, as seen comparing with 
\citep{Blais-Ouellette2001} or with the SPARC mass models \cite{Oh2008}. We want to stress that the differences between 
the IC2574 RC's baryon contribution (gas, stars) in \cite{Kaplinghat2016} from that of SPARC or 
\citep{Blais-Ouellette2001} arise from the different DM properties. 
The halo mass of the simulated galaxy host is $1.8 \times 10^{10}~M_{\odot}$. 
We obtained the virial mass from the stellar mass using \cite{Moster2013}. The stellar mass is 
$M_{\ast}=10^{8.7}~M_{\odot}$. The galaxy effective radius is $R_{\rm eff}=2.8~{\rm kpc}$.

Recently \cite{Santos2017} showed how the RC of IC2574 can be naturally obtained taking into account SN feedback, 
with a similar approach to ours.

\section{Diversity}\label{sect:diversity}
Despite the fact that the RCs of dwarf galaxies are on average cored, individual fits to galaxy RCs show inner slopes 
ranging from $\alpha \simeq 0$ to cusps, while for cored profiles, the central densities can differ by a factor 
of 10 for galaxies inhabiting similar halos \cite{KuziodeNaray2010} and the situation becomes more complicated at 
higher masses. For many objects, \cite{Simon2005} found cored and cuspy profiles in dwarfs which are similar while 
\cite{deBlok2008} observed a tendency to flatter profiles in less massive galaxies.

Such diversity was quantified by \cite{Oman2015} comparing the circular velocity at 2~kpc, $V_{\rm 2kpc}$, with a 
fixed value of the maximum of circular velocity ($V_{\rm max}$). For $50 <V_{\rm max} <250~{\rm km/s}$ there is a 
scatter of a few in $V_{\rm 2kpc}$. \cite{Creasey2017} studied the problem in the SIDM scenario, and found that SIDM 
alone cannot explain the scatter, since the resolution requires baryonic physics must also be taken into account.

In Fig.~\ref{fig:RC}, we plot (left panel) $V_{\rm 2kpc}$ versus the outermost measured circular velocity 
$V_{\rm R last}$. The dashed line is the expectation for a NFW density profile. 
The dots are the observed values from \cite{Lelli2016} and the open squares are the prediction of our model. 
The violet thick line represents the mean trend line. The sample obtained by \cite{Lelli2016} contains galaxies with 
stellar mass in the range $M_{\ast}\simeq 5\times 10^6-10^{11.5}~M_{\odot}$ corresponding to circular velocities in the 
range $15-300~{\rm km/s}$. 
In order to compare their results with our model, we chose simulated galaxies with stellar mass in the range of 
\cite{Lelli2016} and we calculated $V_{\rm 2kpc}$ and $V_{\rm R last}$.
\footnote{$\rm R last$ for simulated galaxies is given from the linear fit to the SPARC galaxies distributed in the 
plane $\rm M_{\ast}-R last$.}

As the dots show, at fixed $V_{\rm R last}$, the scatter in $V_{\rm 2kpc}$ can be as large as a factor of four. 
Such scatter cannot be explained by the $\Lambda$CDM model, as it produces cuspy and self-similar halo density 
profiles, with a single parameter (concentration parameter or halo mass), in contrast to the cores displayed by many 
dwarfs. 
In the $\Lambda$CDM model, the much larger amount of DM in the halo cusp than the baryons ``freezes" the scatter in 
$V_{\rm 2kpc}$, produced, conversely, by the spread in the baryon distribution. 
Baryon physics heats DM and enlarge the galaxies, reducing the inner DM content. As shown in 
\cite{DelPopolo2010,DelPopolo2016b}, the inner slope of the halo density profile is mass dependent. Milky Way-sized 
galaxies tend to have cuspy profiles while dwarf-sized galaxies cored profiles. Ultrafaint dwarf galaxies tend to be 
more cuspy than dwarfs. 
Therefore, the scatter seen in Fig.~\ref{fig:RC} originates from the mass dependence of the core formation process and 
the effects of environment, as described in \cite{DelPopolo2012a}.

Our model successfully recovers the scatter and distribution of the RCs shapes because baryon physics gives rise to 
different responses in the halo of simulated galaxies. 
The right panel of Fig.~\ref{fig:VV} represents the inner slope of the DM halo obtained in \citep{DiCintio2014}, 
dashed line, in the present paper, dot-dashed line, and in \citep{Tollet2016}, solid line. All the curves were obtained 
as in \citep{DiCintio2014} by fitting the DM profile with a power law in the radial range $0.01<r/R_{\rm vir}<0.02$, 
being $R_{\rm vir}$ the virial radius. 
The shaded blue and grey region represents the 1-$\sigma$ scatter in our result and \citep{Tollet2016}.

The plot shows that the core formation mechanism and $\alpha$ are strongly dependent on 
$M_{\ast}/M_{\rm halo}$\footnote{The correlation between $\alpha$ and $M_{\ast}/M_{\rm halo}$ can be expressed in terms 
of $M_{\ast}$, using for example the relation in\cite{Moster2013}.} 
with a minimum value of $\alpha$ at masses $M_{\ast}/M_{\rm halo} \simeq 10^{-2}$ corresponding to 
$M_{\ast} \simeq 10^8 M_{\odot}$ \citep{DiCintio2014,Tollet2016,DelPopolo2016a} due to the maximum effects of baryon 
physics. For smaller masses the profile steepens because of the relative decrease of stars 
(ratio $M_{\ast}/M_{\rm halo}$). Since the profile tends to steepen for $M_{\ast}\lesssim 10^8 M_{\odot}$ and 
$V_{\rm R last}$ is proportional to $M_{\ast}$, we should expect a self-similar behaviour, similar to the NFW RCs, as 
observed in Fig.~\ref{fig:VV}, left panel. 
For $M_{\ast}\geq 10^8 M_{\odot}$, the increase in stellar masses gives rise to a deepening of the potential well and a 
reduction of the effects of baryon physics, with a consequent steepening of the profile. 
Notice that, because we account for star formation modification by AGN feedback which counters baryon cooling, our 
trend line agrees with SPARC at $V_{\rm R last}\ge 150~{\rm km/s}$. 
Moreover, hydrodynamics simulations usually examine isolated galaxies while in our model the account of tidal 
interactions makes it more environment dependent.

\section{Conclusions}\label{sect:conclusions}
The $\Lambda$CDM model exhibits some problems at small scales, and in particular predicts an excess of DM in the 
central parts of galaxies and clusters. In this paper, we showed that a unified solution to the problem can be 
obtained within the $\Lambda$CDM framework without introducing different forms of DM 
\citep[as done in][instead]{Kaplinghat2016}. 
With a semi-analytic model whose main mechanism is based on dynamical friction, we simulated the clusters studied 
by \citep{Newman2013b} and compared the density profiles with those they obtained: those profiles were re-obtained 
correctly by our model.

We displayed one of those density profiles (Fig.~\ref{fig:bgfr}). 

We then simulated a sample of galaxies similar to the SPARC compilation, also containing the galaxies studied by 
\citep{Kaplinghat2016}, finding again a good agreement with data, as shown in the case of one of the most 
complicated galaxy RCs to reproduce, namely that of IC2574 (Fig.~\ref{fig:RC}). 

We want to stress that to match observations we re-simulated the object studied by targeting not only the total 
halo and baryonic masses but also the observed radial density distribution (in the case of Fig.~\ref{fig:bgfr}) and 
rotation curve (in the case of Fig.~\ref{fig:RC}).

Finally, we studied the ``diversity" problem using the simulated galaxies, and comparing their $V_{\rm 2kpc}$, for 
given values of $V_{\rm R last}$ with the compilation in \citep{Lelli2016}. We show that baryon physics gives rise to 
RCs very different from each other, due to the dependence of the RC from their total and stellar mass, together with 
environment. This explains the scatter in the $V_{\rm 2kpc}$-$V_{\rm R last}$ plane.

\section*{Acknowledgements}
F.P. acknowledges support from STFC grant ST/P000649/1. A.D.P.  was supported by the Chinese Academy of Sciences and  
by the President’s International Fellowship Initiative, Grant No. 2017 VMA0044. The authors thank an anonymous referee 
whose comments helped to improve the scientific content of this work.

\bibliographystyle{apsrev4-1}
\bibliography{unified.bbl}

\label{lastpage}

\end{document}